\documentclass[12pt]{article}

\catcode`\@=11
\@addtoreset{equation}{section}

\evensidemargin \oddsidemargin

\usepackage[dvipdfmx]{graphicx,hyperref}
\hypersetup{
setpagesize=false,
bookmarksnumbered=true,
bookmarksopen=true,
colorlinks=true,
linkcolor=black,
citecolor=black,
urlcolor=black,}

\usepackage{mathrsfs,amsbsy,amssymb,latexsym,amsfonts,amsmath,
amssymb,
}

 \usepackage{bm}

\usepackage[dvipdfmx]{graphicx}

\usepackage{color}

\newcommand\CM{\mathcal{M}}

\newcommand\pa{\partial}

\newcommand\bbZ{\mathbb{Z}}

\newcommand\nn{\nonumber}

\newcommand\alphah{{\hat\alpha}}
\newcommand\thetab{{\bm\theta}}

\newcommand\1{\boldsymbol{1}}

\newcommand\adss[2]{AdS$_{#1}\times$S$^{#2}$}

\newcommand\e{{\rm e}}

\renewcommand\d{{\rm d}}

\newcommand\Str{\mathrm{Str} }

\newcommand\bref[1]{(\ref{#1})}

\begin{document}

\vspace*{3cm}

\begin{center}
{\Large \bf 
D-branes
from Pure Spinor Superstring
\\in 
\adss{5}{5}
Background
}
\vspace*{3cm}\\
{\large Sota Hanazawa\footnote{{e-mail: 16nd109n@vc.ibaraki.ac.jp}}
and
Makoto Sakaguchi\footnote{{e-mail: makoto.sakaguchi.phys@vc.ibaraki.ac.jp}}
}
\end{center}
\vspace*{1.5cm}
\begin{center}
Department of Physics,
Ibaraki University,
Mito 310-8512,
Japan
\end{center}

\vspace{3cm}

\begin{abstract}
We examine the surface term for
the BRST transformation of the open pure spinor superstring in an \adss{5}{5}
background.
We find that the boundary condition to eliminate the surface term leads to 
a classification of possible configurations of 1/2 supersymmetric D-branes.

\end{abstract}

\thispagestyle{empty}
\setcounter{page}{0}

\newpage

\tableofcontents

\section{Introduction}

Before
the pure spinor formulation of the superstring
was initiated by Berkovits \cite{PS},
there were mainly two superstring formulations,
a Ramond-Neveu-Schwarz (RNS) formulation and
a Green-Schwarz (GS) formulation.
The RNS superstring
is described by a superconformal field theory on the two-dimensional world-sheet.
So it is difficult to read off the target space geometry  coupling to Ramond-Ramond fields,
because spacetime supersymmetry emerges only after the GSO projection.
On the other hand, for the GS superstring,
spacetime supersymmetry is manifest from the outset.
The action has world-sheet fermionic gauge symmetry,
called $\kappa$-symmetry, instead of world-sheet supersymmetry.
This makes it  difficult 
to covariantly quantize the GS  superstring even in flat spacetime.
In the pure spinor superstring\footnote{
The extended versions of the pure spinor superstring
were proposed
in \cite{without,without 2}
which introduced new ghosts to relax the pure spinor constraint,
and
in \cite{doubled}
which  introduced
doubled spinor degrees of freedom
with a compensating local supersymmetry.},
the $\kappa$-symmetry in the GS  superstring is replaced with the BRST symmetry.
The pure spinor superstring
can be quantized in a super-Poincar\'e covariant manner.

Furthermore,
the pure spinor superstring in an \adss{5}{5} background 
with Ramond-Ramond flux
\cite{PS}
is shown to be consistent even at the quantum level
\cite{PS quantum,PS one-loop}.
The 
action
is composed of $\mathfrak{psu}(2,2|4)$ currents $J$, 
and the left- and right-moving ghosts, $(\lambda^\alpha,w_\alpha)$,
and  $(\hat\lambda^\alphah,\hat w_\alphah)$, respectively.
The ghosts 
satisfy the pure spinor constraints
$\lambda\gamma^A\lambda=\hat\lambda\gamma^A\hat\lambda=0$
($A=0,1,\cdots,9$).
The pure spinor superstring in the \adss{5}{5} background,
as well as the GS superstring in the \adss{5}{5} background given in  \cite{GS AdS5xS5},
is integrable in the sense that
infinitely many conserved charges are constructed
\cite{BPR,PS AdS integrable}
(see also \cite{integrability GS RS}).
Nevertheless, for the detailed study of the AdS/CFT correspondence
\cite{Maldacena}, 
covariant quantization of the superstring should be useful.
Though the action of the pure spinor superstring in the \adss{5}{5} background is bilinear in the current $J$,
its quantization is still difficult
because the $J$ is not (anti-)holomorphic
unlike the principal chiral model.
We need more effort to quantize  the pure spinor superstring covariantly.

\medskip
The purpose of this paper is to study D-branes in the \adss{5}{5} background.
A D-brane
is a solitonic object in string theory,
and is characterized by the Dirichlet boundary condition of an open string.
The classical BRST invariance of the open pure spinor superstring
in a  background
implies that the background fields satisfy full non-linear
 equations of motion for a supersymmetric Born-Infeld action
\cite{PS open background}.
This is the open string version of \cite{PS closed background}
in which  the classical BRST invariance of the closed pure spinor superstring
in a curved background
was shown to imply that the background fields satisfy 
full non-linear equations of motion
for the type-II supergravity.
For D-branes in  the \adss{5}{5} background,
supersymmetric D-brane configurations are derived
in  \cite{Dp AdS pp}
by examining equations of motion
 for a Dirac-Born-Infeld action
for each D-brane embedding ansatz.

\medskip

In the present paper, 
we will examine D-branes in the \adss{5}{5} background by using the open pure spinor superstring.
Especially, we concentrate ourselves on the BRST invariance in the presence of the boundary.
Namely, we examine the surface term for
the BRST transformation of the open pure spinor superstring in the \adss{5}{5} background.
We will find that the boundary condition to eliminate the surface term leads to 
a classification of possible configurations of 1/2 supersymmetric D-branes.
This approach is the pure spinor superstring version of \cite{GS kappa 1,GS kappa 2}\footnote{
A covariant approach to study D-branes in flat spacetime was proposed by Lambert and West in \cite{LW}. 
}.
In \cite{GS kappa 1,GS kappa 2}, the boundary condition for the $\kappa$-symmetry surface term
of the GS superstring in the \adss{5}{5} background
 was shown to lead to a classification of possible configurations of  1/2 supersymmetric D-branes.
 We find that our result is consistent with those obtained in \cite{Dp AdS pp,GS kappa 1,GS kappa 2}. 
One of the main advantages in our approach is that 
the derivation is much simpler than the one by using the Dirac-Born-Infeld  action
 and the GS superstring action.
This is because the pure spinor superstring action is bilinear in the currents,
and because  we don't need to deal with the $\kappa$-symmetry
variation which is highly non-linear.

\medskip

This paper is organized as follows.
In section 2, after introducing the pure spinor superstring in the \adss{5}{5} background,
we examine the BRST invariance of the open superstring action
and extract the surface term.
For the BRST invariance to be preserved even in the presence of the boundary,
the surface term must be eliminated by a certain boundary condition.
In section 3, we fix the boundary conditions by examining a few terms contained in the surface term.
The boundary conditions lead us to a classification of possible 1/2 supersymmetric D-branes in the \adss{5}{5} background.
In section 4, the boundary conditions fixed above
are shown to eliminate all terms contained in the surface term.
The last section is devoted to a summary and discussions.
Our notation and convention are summarized in Appendix.

\section{Pure Spinor Superstring in \adss{5}{5} background} 

The manifestly covariant action of the pure spinor superstring in the \adss{5}{5} background 
\cite{PS,PS quantum,AdS PS action}
(see e.g. \cite{review} for reviews)
is composed of three parts
\begin{eqnarray}
S&=&S_\sigma+S_\mathrm{WZ}+S_\mathrm{gh}~,
\end{eqnarray}
with
\begin{eqnarray}
S_\sigma&=&\frac{1}{2}\left\langle
J_2\bar J_2 +J_1\bar J_3+J_3\bar J_1
\right\rangle~,
\\
S_\mathrm{WZ}&=&\frac{1}{4}\left\langle
J_3\bar J_1 - J_1 \bar J_3
\right\rangle~,
\\
S_\mathrm{gh}&=&\left\langle
w\bar\pa \lambda
+\hat w \pa \hat \lambda
+N\bar J_0
+\hat N J_0
-N\hat N
\right\rangle~,\end{eqnarray}
where $\langle \cdots \rangle$ stands for 
$\frac{1}{\pi\alpha'}\int \d^2\sigma\, \Str(\cdots)$.
Here
$J=\frac{1}{2}({J}_\tau+ {J}_\sigma)$
and $\bar J=\frac{1}{2}({J}_\tau- {J}_\sigma)$, 
 are the left- and right-moving currents,
respectively.
The
 $J_\xi$ ($\xi=\tau,\sigma$)
stands for the pullback of the Cartan one-form on the worldsheet,
$J_\xi=g^{-1}\pa_\xi g$,
with 
$g\in$ PSU(2,2$|$4)/(SO(1,4)$\times$ SO(5)).
Furthermore $J_\xi$ is decomposed into four parts,
under the $\bbZ_4$-graded decomposition of 
$\mathfrak{psu}(2,2|4)$, namely
\begin{eqnarray}
J_\xi&=&J_{0\xi}+J_{1\xi}+J_{2\xi}+J_{3\xi},\nn\\
J_{0\xi}&=&\frac{1}{2}{J}_{0\xi}^{ab}M_{ab}+\frac{1}{2}{J}_{0\xi}^{a'b'}M_{a'b'}~,~
J_{1\xi}=q_\alpha{J}_{1\xi}^\alpha ~,~
J_{2\xi}={J}_{2\xi}^aP_a+{J}_{2\xi}^{a'}P_{a'}~,~
J_{3\xi}={\hat q_{\hat\alpha}J}_{3\xi}^{\hat\alpha}
\label{J}\,.~~
\end{eqnarray}
The set of generators $\{M_{ab},M_{a'b'},P_{a},P_{a'},q_\alpha, \hat q_{\hat\alpha}\}$
of $\mathfrak{psu}(2,2|4)$
satisfies (anti-)commutation relations given in \bref{psu bosonic}
and \bref{psu fermionic 16}.
The pure spinor variables are defined as
\begin{eqnarray}
\lambda=\lambda^\alpha q_\alpha~,~~
\hat\lambda=\hat\lambda^\alphah \hat q_\alphah~,~~
w=w_\alpha (\tilde \gamma^{0\cdots4})^{\alpha\alphah}\hat q_\alphah~,~~
\hat w=\hat w_\alphah (\tilde \gamma^{0\cdots4})^{\alphah\alpha}q_\alpha~,
\end{eqnarray}
where $(\lambda^\alpha,w_\alpha)$ and $(\hat\lambda^\alphah, \hat w_\alphah)$ 
are left- and right-moving ghosts, respectively.
In terms of these ghosts the Lorentz currents are given as
$N=-\{ w,\lambda\}$
and $ \hat N=-\{\hat w,\hat \lambda\}$.

\subsection{BRST invariance}

The BRST transformation of the action is examined below.
We will not drop any surface term here.
In the next section we will consider the boundary condition for the surface term to be eliminated,
and show that the condition leads us to a classification of  possible configurations of 1/2 supersymmetric D-branes
in the \adss{5}{5} background.

First we examine $S_\sigma$.
The BRST transformation law of currents
with a Grassmann odd parameter  $\varepsilon$
is given as \cite{PS quantum}
\begin{eqnarray}
\varepsilon Q(J_1)&=&
\nabla(\varepsilon \lambda)
+[J_2,\varepsilon \hat\lambda]~,~~~
\varepsilon Q(\bar J_1)=
\bar\nabla(\varepsilon \lambda)
+[\bar J_2,\varepsilon \hat\lambda]~,\nn\\
\varepsilon Q(J_2)&=&[J_1,\varepsilon \lambda]+[J_3,\varepsilon \hat\lambda]~,~~~
\varepsilon Q(\bar J_2)=[\bar J_1,\varepsilon \lambda]+[\bar J_3,\varepsilon \hat\lambda]~,\nn\\
\varepsilon Q(J_3)&=&
\nabla(\varepsilon \hat \lambda)
+[J_2,\varepsilon \lambda]~,~~~
\varepsilon Q(\bar J_3)=
\bar \nabla(\varepsilon \hat \lambda)
+[\bar J_2,\varepsilon \lambda]~,\nn\\
\varepsilon Q(J_0)&=&[J_3,\varepsilon \lambda]+[J_1,\varepsilon \hat\lambda]~,~~~
\varepsilon Q(\bar J_0)=[\bar J_3,\varepsilon \lambda]+[\bar J_1,\varepsilon \hat\lambda]~,
\label{BRST J}
\end{eqnarray}
where $\nabla A \equiv \pa A+[J_0,A]$
and  $\bar \nabla A \equiv \bar\pa A+[\bar J_0,A]$.
By using \bref{BRST J}, we obtain
\begin{eqnarray}
\varepsilon Q(S_\sigma)&=&\frac{1}{2}\left\langle
J_1\bar\nabla(\varepsilon \hat\lambda)
+\nabla(\varepsilon \lambda) \bar J_3
+J_3\bar\nabla(\varepsilon\lambda)
+\nabla(\varepsilon \hat \lambda) \bar J_1
\right\rangle~.
\label{BRST S_sigma}
\end{eqnarray}
To derive this expression we have used the cyclicity of the $\Str$,
for example $\Str(J_2[\bar J_1,\varepsilon\lambda])=\Str( \varepsilon\lambda [J_2,\bar J_1])$.

Next we consider $S_\mathrm{WZ}$.
Using \bref{BRST J}, one derives
\begin{eqnarray}
\varepsilon Q(S_\mathrm{WZ})&=&
\frac{1}{4}
\left\langle
\nabla(\varepsilon \hat\lambda)\bar J_1
+J_3\bar \nabla(\varepsilon\lambda) 
-\nabla(\varepsilon\lambda)\bar J_3
-J_1\bar \nabla(\varepsilon\hat \lambda)
\right.
\nn\\
&& \left. ~~+\varepsilon\lambda([\bar J_1,J_2]-[ J_1,\bar J_2])
+\varepsilon\hat \lambda([J_3,\bar J_2]-[ \bar J_3, J_2])
\right\rangle~.
\label{BRST S_W'}
\end{eqnarray}
By using
Maurer-Cartan equations 
\begin{eqnarray}
\nabla \bar J_3-\bar\nabla J_3=[\bar J_1,J_2]-[ J_1,\bar J_2]~,~~
\nabla \bar J_1-\bar\nabla J_1=[\bar J_3,J_2]-[ J_3,\bar J_2]~,
\end{eqnarray}
the second line of the right-hand side of \bref{BRST S_W'}
may be rewritten as
\begin{eqnarray}
\varepsilon Q(S_\mathrm{WZ})&=&
\frac{1}{4}
\left\langle
\nabla(\varepsilon \hat\lambda)\bar J_1
+J_3\bar \nabla(\varepsilon\lambda) 
-\nabla(\varepsilon\lambda)\bar J_3
-J_1\bar \nabla(\varepsilon\hat \lambda)
\right.
\nn\\
&& \left. ~~+\varepsilon\lambda(\nabla \bar J_3-\bar\nabla J_3)
-\varepsilon\hat \lambda(\nabla \bar J_1-\bar\nabla J_1)
\right\rangle~.
\label{BRST S_W}
\end{eqnarray}

Finally we examine $S_\mathrm{gh}$.
The BRST transformation law of ghosts
\begin{eqnarray}
\varepsilon Q(w)=-J_3\varepsilon~,~~
\varepsilon Q(\hat w)=-\bar J_1\varepsilon~,~~
\varepsilon Q(\lambda)=\varepsilon Q(\hat \lambda)=0
\end{eqnarray}
implies that
\begin{eqnarray}
\varepsilon Q(N)=[J_3,\varepsilon\lambda]~,~~
\varepsilon Q(\hat N)=[\bar J_1,\varepsilon\hat \lambda]~.
\end{eqnarray}
Further noting that
\begin{eqnarray}
\Str(N[\bar J_3,\varepsilon\lambda])
=\Str(-\bar J_3\varepsilon[\lambda\lambda,w])=0~,~~
\Str(\hat N[J_1,\varepsilon\hat \lambda])
=\Str(-J_1\varepsilon[\hat\lambda\hat\lambda,\hat w])=0~,~~
\end{eqnarray}
which follow from the pure spinor conditions $\{\lambda,\lambda\}=\{\hat\lambda,\hat\lambda\}=0$,
we can derive
\begin{eqnarray}
\varepsilon Q(S_\mathrm{gh})=
\left\langle
-J_3 \bar \nabla (\varepsilon \lambda)
-\bar J_1 \nabla (\varepsilon \hat\lambda)
\right\rangle~.
\label{BRST S_gh}
\end{eqnarray}

Gathering all results obtained above together, 
we find that the BRST transformation of $S$ is
\begin{eqnarray}
\varepsilon Q(S)&=&\frac{1}{4}\left\langle
\nabla(\varepsilon \lambda \bar J_3 -\varepsilon \hat\lambda \bar J_1)
-\bar\nabla(\varepsilon \lambda J_3 - \varepsilon\hat\lambda  J_1)
\right\rangle\nn\\
&=&\frac{1}{4}\left\langle
\pa(\varepsilon \lambda \bar J_3 -\varepsilon \hat\lambda \bar J_1)
-\bar\pa(\varepsilon \lambda J_3 - \varepsilon\hat\lambda  J_1)
\right\rangle~.
\label{surface term}
\end{eqnarray}
In the second equality,
we have used the fact that
$\Str([J_0,\varepsilon\lambda \bar J_3])=0$
and the similar relations.
 
We can conclude that $S$ is BRST invariant as long as this surface term vanishes.
For a closed string, the surface term  always vanishes.
For an open string, however, appropriate boundary conditions are required.
In the next section we will examine these boundary conditions.

\section{Boundary BRST invariance
to D-brane configurations
}

In this section we will examine boundary conditions for the surface term to be eliminated,
and show that they lead us to a classification of possible
1/2 supersymmetric D-brane configurations in the \adss{5}{5} background.

The surface term \bref{surface term}
turns to
\footnote{
$\pa \bar J-\bar\pa J=\pa_\tau J_\sigma +\pa_\sigma J_\tau$, as $\pa=\pa_\tau+\pa_\sigma$ and $\bar\pa=\pa_\tau-\pa_\sigma$.
}
\begin{eqnarray}
\varepsilon Q(S)&=&\frac{1}{4\pi \alpha'}\int \d^2 \sigma\, \Str\left[
\pa_\sigma (\varepsilon \lambda J_3 -\varepsilon \hat\lambda J_1)_\tau
+\pa_\tau (\varepsilon \lambda J_3 -\varepsilon \hat\lambda J_1)_\sigma
\right]\nn\\
&=&\frac{1}{4\pi \alpha'}\int \d\tau\, \Str\left[
(\varepsilon \lambda J_3 -\varepsilon \hat\lambda J_1)_\tau
\right]\Big|_{\sigma=\sigma_*}
\label{surface 2}
\end{eqnarray}
where
we have assumed that the surface term at $\tau=\pm \infty$ vanishes as usual.
The open string boundaries are at $\sigma=\sigma_*$
with $\sigma_*=0,\pi$.
As seen in \bref{J 16},
$J_{1\tau}$ and $J_{3\tau}$ correspond to $q_\alpha L^{1\alpha}_\tau$
and $\hat q_{\hat \alpha} L^{2\hat\alpha}_\tau$, respectively,
where the corresponding Cartan one-form $L^I$ is given in \bref{Cartan J^I}.
It follows that
$J_{1\tau}$ and $J_{3\tau}$ are polynomials in $\theta^I$.
We should note that the surface terms do not cancel out each other.
So we may examine each surface term separately without loss of generality.
Our strategy is as follows. 
First we examine a few
terms
contained in $J_1$ and
$J_3$,
  and fix the boundary condition.
Next we will show that the boundary condition
we have fixed would eliminate 
all terms
in \bref{surface 2}.

First, we shall fix the boundary condition
by examining the following three  terms contained in 
$J_{1\tau}$ and
those in $J_{3\tau}$
\begin{eqnarray}
J_{1\tau}&=& q_\alpha L^{1\alpha}_\tau
= q_\alpha\left(
\pa_\tau\theta+
\frac{1}{2} e^A_\tau \tilde \gamma^{0\cdots 4}\gamma_A\hat\theta
-\frac{i}{6}\tilde \gamma^{0\cdots 4}\gamma_A\hat \theta
\, (\theta\gamma^A\pa_\tau \theta+\hat \theta\gamma^A\pa_\tau\hat\theta)
\right)
+\cdots~,
\label{J_1 tau}\\
J_{3\tau}&=& \hat q_\alphah L^{2\alphah}_\tau
= \hat q_\alphah \left(
\pa_\tau\hat \theta
-\frac{1}{2} e^A_\tau \tilde \gamma^{0\cdots 4}\gamma_A\theta
+\frac{i}{6}\tilde \gamma^{0\cdots 4}\gamma_A\theta\, 
(\theta\gamma^A\pa_\tau \theta+\hat \theta\gamma^A\pa_\tau\hat\theta)
\right)
+\cdots.
\label{J_3 tau}
\end{eqnarray}
The first two terms in the most right-hand sides 
in the above equations
come from $D\theta$ defined in \bref{D thetra},
while the last one is contained in $\frac{m^2}{3!}D\theta$
where $m^2$ is defined in \bref{m^2}.
After fixing the boundary condition we shall show that  the boundary condition eliminates
all surface terms in section 4. 

Substituting \bref{J_1 tau} and \bref{J_3 tau} into  \bref{surface 2},
we obtain the corresponding surface terms,
which we shall denote as $\delta_0S$,
\begin{eqnarray}
\delta_0 S
&=&\frac{1}{4\pi \alpha'}\int \d\tau\, 
\varepsilon\Bigg[
\lambda\gamma^{0\cdots 4}
\left(
\pa_\tau\hat \theta
-\frac{1}{2} e^A_\tau \tilde \gamma^{0\cdots 4}\gamma_A\theta
+\frac{i}{6}\tilde \gamma^{0\cdots 4}\gamma_A\theta\, 
(\theta\gamma^A\pa_\tau \theta+\hat \theta\gamma^A\pa_\tau\hat\theta)
\right)\nn\\
&&\hspace{10mm}
+\hat\lambda \gamma^{0\cdots 4}
\left(
\pa_\tau\theta+
\frac{1}{2} e^A_\tau \tilde \gamma^{0\cdots 4}\gamma_A\hat\theta
-\frac{i}{6}\tilde \gamma^{0\cdots 4}\gamma_A\hat \theta
\, (\theta\gamma^A\pa_\tau \theta+\hat \theta\gamma^A\pa_\tau\hat\theta)
\right)
\Bigg]\Bigg|_{\sigma=\sigma_*}
\label{delta 0 S 16}
\end{eqnarray}
where 
we have used 
$\Str(q_\alpha \hat q_{\hat\beta})=(\gamma^{0\cdots 4})_{\alpha\hat\beta}$ 
and 
$\Str(\hat q_\alphah q_{\beta})=-(\gamma^{0\cdots 4})_{\alphah\beta}$.
For our purpose,
it is convenient to rewrite \bref{delta 0 S 16} in a 32-component notation.
Defining 
\begin{eqnarray}
\bm \lambda^I\equiv \left(\begin{array}{c}\lambda^I \\0\end{array}\right)~,~~~
\bm\theta^I \equiv \left(\begin{array}{c}\theta^I \\0\end{array}\right)~,
\end{eqnarray}
where $(\lambda^1,\lambda^2)=(\lambda,\hat\lambda)$
and $(\theta^1,\theta^2)=(\theta,\hat\theta)$,
and using the 32 component spinor notation
given in Appendix A,
we find that \bref{delta 0 S 16} 
may be simplified to
\begin{eqnarray}
\delta_0 S
=\frac{1}{4\pi \alpha'}\int \d\tau\, 
\varepsilon\left[
\bar{\bm\lambda} I \sigma_1 \pa_\tau \bm\theta
+\frac{1}{2} e^A_\tau \bar{\bm\lambda}\Gamma_A\sigma_3\bm\theta
-\frac{i}{6} \bar{\bm\lambda} \Gamma_A \sigma_3 \bm\theta\,
\bar{\bm\theta}\Gamma^A \pa_\tau \bm\theta
\right]\Bigg|_{\sigma=\sigma_*}
~.
\label{surface terms}
\end{eqnarray}
To show this,  the following relations are useful
\begin{eqnarray}
\bar{\bm\lambda}^I I (\sigma_1)_{IJ}\pa\bm\theta^J
&=& \lambda^1\gamma^{0\cdots 4} \pa\theta^2
+ \lambda^2\gamma^{0\cdots 4} \pa\theta^1~,\\
\bar{\bm\lambda}^I \Gamma_A  (\sigma_3)_{IJ}\bm\theta^J
&=& \lambda^1\gamma_A \theta^1
- 
\lambda^2\gamma_A \theta^2~,\\
\bar{\bm\theta}\Gamma^A \pa_\tau\bm\theta
&=&
\theta^1\gamma^A \pa_\tau \theta^1
+\theta^2\gamma^A \pa_\tau \theta^2~.
\end{eqnarray}

For bosonic coordinates, we impose the  boundary conditions as follows:
Neumann boundary condition 
$e^{\bar A}_\sigma =\pa_\sigma x^\mu e_\mu^{\bar A}=0$
for $\bar A=\bar A_0,\cdots,\bar A_{p}$,
or Dirichlet boundary condition
$e^{\underline A}_\tau=\pa_\tau x^\mu e^{\underline A}_\mu=0$
for $\underline A=\underline A_{p+1},\cdots,\underline  A_{9}$.
This boundary condition eliminates the surface term $\delta x^\mu e_\mu^{A} e_\sigma^B \eta_{AB}$
at $\sigma=\sigma_*$.
In order to delete $\delta_0 S$, we must impose boundary conditions on
$\bm\theta$ and $\bm\lambda$.
The boundary condition we shall  impose on $\bm\theta$ is
\begin{eqnarray}
\bm\theta=M\, \bm\theta~,~~~
M=s\Gamma^{\bar A_0 \cdots \bar A_{p}}\otimes \rho
\label{BC}
\end{eqnarray}
where $\rho_{IJ} \in \{ 1,\sigma_1, i\sigma_2,\sigma_3\}$ is a two-by-two matrix  acting on $\bm\theta^I$ ($I=1,2$).
For the reality of $\bm\theta$, we choose $s$ as $s=\pm 1$.
The boundary condition leads to 1/2 supersymmetric D-branes.
As $\bm\theta^I$ are a pair of  Majorana-Weyl spinors satisfying $\bm\theta=\Gamma_{11}\bm\theta$,  
we find  $p=$ odd for consistency, $[M,\Gamma_{11}]=0$.
For the boundary condition on $\bm\lambda$, we must impose $\bm\lambda =M\bm\lambda$.
This is necessary for the BRST transformation to be non-trivial even at the boundary.
In fact, the BRST transformation of $J_1$ in  \bref{BRST J},
namely  
$
\varepsilon Q (\pa \theta^\alpha) = \varepsilon \pa \lambda^\alpha +\cdots ~,
$
is consistent if we impose the same boundary condition on $\bm\theta$ and $\bm\lambda$.

We shall examine each term contained in \bref{surface terms} below
so that we will fix $p$ and $\rho$.
Let us begin with 
examining the second term in the right hand side  of \bref{surface terms}.
Because $e^{\underline A}_\tau=0$,
\begin{eqnarray}
 \bar{\bm\lambda}\Gamma_{\bar A}\sigma_3\bm\theta=0
 \label{lambda theta}
\end{eqnarray}
must be satisfied.
It follows from \bref{lambda theta}
that in order to delete  the third term in the right hand side  of \bref{surface terms},
\begin{eqnarray}
\bar{\bm\theta}\Gamma^{\underline A} \pa_\tau \bm\theta=0~
\label{theta theta}
\end{eqnarray}
must be satisfied.
We examine \bref{theta theta} first.
Noting that 
$CM=\mp \alpha M^TC$ with $\rho=\alpha \rho^T$
for $p=\{{1\atop 3} $ mod 4,
respectively,
we derive 
\begin{eqnarray}
\bar{\bm\theta}\Gamma^{\underline A} \pa_\tau  \bm\theta
=\bar{\bm\theta} \Gamma^{\underline A} M\pa_\tau  \bm\theta
={\bm\theta}^T CM  \Gamma^{\underline A} \pa_\tau  \bm\theta
=\mp \alpha \overline{M\bm\theta}  \Gamma^{\underline A} \pa_\tau  \bm\theta
=\mp \alpha \bar{\bm\theta}  \Gamma^{\underline A} \pa_\tau  \bm\theta~,
\label{tt cal}
\end{eqnarray}
so that $\alpha$ is fixed as $\alpha =\pm 1$ for \bref{theta theta}.
It means that $\rho = \pm \rho^T$ for  $p=\{{1\atop 3} $ mod 4.

Now, we return to \bref{lambda theta}.
We derive, defining $\beta$ by $\sigma_3\rho = \beta \rho\sigma_3$,
\begin{eqnarray}
\bar{\bm\lambda}\Gamma_{\bar A}\sigma_3\bm\theta
=\bar{\bm\lambda}\Gamma_{\bar A}\sigma_3M\bm\theta
=\beta \bar{\bm\lambda}\Gamma_{\bar A}M\sigma_3\bm\theta
=-\beta \bar{\bm\lambda}M\Gamma_{\bar A}\sigma_3\bm\theta
=\beta \overline{M\bm\lambda}\Gamma_{\bar A}\sigma_3\bm\theta
=\beta \bar{\bm\lambda}\Gamma_{\bar A}\sigma_3\bm\theta~.~~
\end{eqnarray}
It implies that $\beta$ is fixed as  $\beta =-1$ for  \bref{lambda theta}.
This means that  $\rho = \sigma_1$ or $i\sigma_2$.
Combining this with the result obtained from  \bref{lambda theta},
we can conclude that $\rho=\sigma_1$ for $p=1$ mod 4, and that
$\rho= i\sigma_2$ for $p=3$ mod 4.
For consistency we require that $M^2=1$.
This implies that the time direction $0$ is a Neumann direction
since $s^2=1$.
The results so far 
coincide with the boundary condition for 1/2 supersymmetric D-branes in flat spacetime.

Finally, we examine  the first term in the right hand side  of \bref{surface terms}
which leads to the additional condition specific to the \adss{5}{5} background.
One may show that
\begin{eqnarray}
\bar{\bm\lambda} I \sigma_1 \pa_\tau \bm\theta
=\bar{\bm\lambda} I \sigma_1 M\pa_\tau \bm\theta
=\pm\bar{\bm\lambda} I M\sigma_1 \pa_\tau \bm\theta
=\pm(-1)^n \bar{\bm\lambda} M I \sigma_1 \pa_\tau \bm\theta
=\mp(-1)^n \bar{\bm\lambda}  I \sigma_1 \pa_\tau \bm\theta~.
\label{n cal}
\end{eqnarray}
In the second equality we have used  $\sigma_1 \rho=\pm \rho \sigma_1 $ for  $p=\{{1\atop 3} $ mod 4.
The third equality follows from $IM=(-1)^nMI$ where $n$ is the number of Neumann directions
contained  in AdS$_5$ spanned by \{0,1,2,3,4\}.
As a result, for $\bar{\bm\lambda} I \sigma_1 \pa_\tau \bm\theta=0$ we must impose
$n=$ even for $p=1$ mod 4 and
$n=$ odd for $p=3$ mod 4.
We summarize the result in  the Table 1
where $(n,n')$ means a D-brane of which world-volume is extended along \adss{n}{n'}.
This gives a classification of 1/2 supersymmetric 
D-brane configurations in the \adss{5}{5} background.
This result is consistent with the ones obtained by using 
the $\kappa$-symmetry variation of the Green-Schwarz superstring
\cite{GS kappa 1,GS kappa 2} 
and by examining D$p$-brane field equations \cite{Dp AdS pp}.

\begin{table}[htp]
\begin{center}
$
\begin{array}{|c|c|c|c|c|c|c|}
\hline p &-1&  1 & 3 & 5 & 7 & 9 \\\hline \mbox{D-brane} &-& (2,0) & (1,3), (3,1) & (2,4), (4,2) & (3,5), (5,3)&- \\\hline 
\end{array}
$
\end{center}
\caption{1/2 supersymmetric D-brane configurations in \adss{5}{5} }
\label{default}
\end{table}

\bigskip

Summarizing the results,
we find that the surface term \bref{surface terms}
vanishes if we impose the boundary conditions
\begin{eqnarray}
\bm\theta=M\bm\theta~,~~~
\bm\lambda=M\bm\lambda~,
\label{bc 1}
\end{eqnarray}
with 
\begin{eqnarray}
M=\left\{\begin{array}{c} s\Gamma^{\bar A_0\cdots \bar A_p}\otimes \sigma_1\\
s\Gamma^{\bar A_0\cdots \bar A_p}\otimes i\sigma_2\end{array}\right.
,~~n=\left\{\begin{array}{c}\mbox{even} \\ \mbox{odd}\end{array}\right. 
\mbox{~~~for~~} p=  
\left\{\begin{array}{c}1 \\3\end{array}\right. \mbox{~mod~} 4~,~~s=\pm 1
\label{bc 2}
\end{eqnarray}
respectively.

\section{Proof of validity in eliminating all surface terms}

In the previous section, we have derived the boundary conditions
\bref{bc 1} with \bref{bc 2}
which eliminate a certain terms
\bref{surface terms}
contained in the surface term \bref{surface 2}.
In this section, we shall show that the boundary conditions fixed in the previous section
may eliminate all terms contained in the surface term \bref{surface 2}.

For this purpose, it is convenient to rewrite the surface term \bref{surface 2}
in the 32-component notation as
\begin{eqnarray}
\varepsilon Q(S)&=&
-\frac{\varepsilon}{4\pi\alpha'}\int\! \d \tau
\left[
\bar{\bm\lambda} I\sigma_1 \bm L_\tau
\right]\big|_{\sigma=\sigma_*}
\label{full surface term}
\end{eqnarray}
where the corresponding Cartan one-form $\bm L$ is given in \bref{J^I 32}.
We will show that the validity of the boundary conditions
for $p=1$ mod 4 and for $p=3$ mod 4,
in turn.

\subsection{$p=1$ mod 4}
We shall show that  the surface term \bref{full surface term}
is eliminated by the boundary conditions
for $p=1$ mod 4:
$\thetab = P_+ \thetab$
and
$\bm\lambda = P_+ \bm\lambda$
where  $P_+=\frac{1}{2}(1+M)$ with 
$M=s\Gamma^{\bar A_0\cdots \bar A_p}\otimes \sigma_1$
and $n=$ even.

First we examine $D_\tau \thetab$
defined in \bref{D theta 32}.
It follows from
$\thetab = P_+ \thetab$
that
\begin{eqnarray}
D_\tau\thetab = P_+ D_\tau\thetab~.
\label{D theta 1}
\end{eqnarray}
In order to derive this relation,
we have used
\begin{eqnarray}
\pa_\tau\thetab&=& P_+ \pa_\tau\thetab~,~~\\
\frac{1}{2}e^{\bar A}_\tau I\Gamma_{\bar A} \epsilon \thetab
&=&\frac{1}{2}e^{\bar A}_\tau I\Gamma_{\bar A} \epsilon P_+ \thetab
=\frac{1}{2}e^{\bar A}_\tau I\Gamma_{\bar A} P_-\epsilon  \thetab
=\frac{1}{2}e^{\bar A}_\tau I P_+\Gamma_{\bar A}\epsilon  \thetab
=P_+\frac{1}{2}e^{\bar A}_\tau I \Gamma_{\bar A}\epsilon  \thetab ~,\\
\frac{1}{4} w^{A B}_\tau \Gamma_{A B}\thetab
&=&\frac{1}{4} w^{\bar A\bar B}_\tau \Gamma_{\bar A\bar B}P_+\thetab
+\frac{1}{4} w^{\underline A\underline B}_\tau \Gamma_{\underline A\underline B}P_+\thetab
=P_+
\left(\frac{1}{4} w^{\bar A\bar B}_\tau \Gamma_{\bar A\bar B}\thetab
+\frac{1}{4} w^{\underline A\underline B}_\tau \Gamma_{\underline A\underline B}\thetab
\right).~~~~
\end{eqnarray}
Here we have assumed that $w^{\bar A \underline B}=0$.
This is because
a D-brane breaks rotational invariance in the plane
spanned by
one of Neumann directions and 
one of Dirichlet directions.

Next we will examine  $\CM^2 P_+$
where $\CM^2$ is defined in \bref{M^2 32}.
Noting that $CP_\pm=P_\mp^T C$
one derives
\begin{eqnarray}
-iI\Gamma_A \epsilon \thetab\, \bar \thetab \Gamma^{A}P_+
&=&
-iI\Gamma_{\bar A} \epsilon \thetab\, \bar \thetab P_- \Gamma^{\bar A}P_+
=-iP_+IP_+\Gamma_{\bar A}P_- \epsilon P_+ \thetab\, \bar \thetab P_- \Gamma^{\bar A}P_+~,\\
\frac{i}{2} \Gamma_{AB} \thetab\,\bar\thetab \hat \Gamma^{AB} \epsilon P_+
&=&
\frac{i}{2} \Gamma_{AB} \thetab\,\bar\thetab \hat \Gamma^{AB} P_-\epsilon P_+
\nn\\
&=&
\frac{i}{2} \Gamma_{\bar A\bar B} \thetab\,\bar\thetab P_- \hat \Gamma^{\bar A\bar B}  P_-\epsilon P_+
+\frac{i}{2} \Gamma_{\underline A \underline B} \thetab\,\bar\thetab P_- \hat \Gamma^{\underline A \underline B} P_-\epsilon P_+
\nn\\
&=&
P_+\frac{i}{2} \Gamma_{\bar A\bar B} \thetab\,\bar\thetab P_- \hat \Gamma^{\bar A\bar B}  P_-\epsilon P_+
+P_+\frac{i}{2} \Gamma_{\underline A \underline B} \thetab\,\bar\thetab P_- \hat \Gamma^{\underline A \underline B} P_-
\epsilon P_+~.
\end{eqnarray}
It follows that
\begin{eqnarray}
\CM^2P_+ = P_+ \CM^2 P_+~.
\label{CM^2 1}
\end{eqnarray}
Gathering the results \bref{D theta 1} and \bref{CM^2 1} together we  obtain
$\bm L_\tau = P_+ \bm L_\tau$.
Using this we may derive
\begin{eqnarray}
\bar{\bm\lambda} I\sigma_1 \bm L_\tau
&=&
\bar{\bm\lambda} I\sigma_1 P_+\bm L_\tau
=\bar{\bm\lambda} P_+ IP_+\sigma_1 P_+\bm L_\tau
=\overline {P_- \bm\lambda} P_+ IP_+\sigma_1 P_+\bm L_\tau=0~.
\end{eqnarray}
This shows that the boundary conditions for $p=1$ mod 4
eliminate the surface term \bref{full surface term}.

\subsection{$p=3$ mod 4}
We show that
the surface term \bref{full surface term}
is eliminated by
the boundary conditions
for $p=3$ mod 4:
$\thetab = P_+ \thetab$
and
$\bm\lambda = P_+ \bm\lambda$
where  $P_+=\frac{1}{2}(1+M)$ with 
$M=\Gamma^{\bar A_0\cdots \bar A_p}\otimes i\sigma_2$
and $n=$ odd.

First we examine $D_\tau \thetab$.
The calculation similar  to the one in  the case with $p=1$ mod 4,
 except for 
\begin{eqnarray}
\frac{1}{2}e^{\bar A}_\tau I\Gamma_{\bar A} \epsilon \thetab
&=&\frac{1}{2}e^{\bar A}_\tau I\Gamma_{\bar A} \epsilon P_+ \thetab 
=\frac{1}{2}e^{\bar A}_\tau I\Gamma_{\bar A} P_+\epsilon  \thetab 
=\frac{1}{2}e^{\bar A}_\tau I P_-\Gamma_{\bar A}\epsilon  \thetab 
=P_+\frac{1}{2}e^{\bar A}_\tau I \Gamma_{\bar A}\epsilon  \thetab ~,
\end{eqnarray}
leads us to 
\begin{eqnarray}
D_\tau\thetab = P_+ D_\tau\thetab~.
\label{D theta 2}
\end{eqnarray}
Next we will examine  $\CM^2 P_+$.
Noting that $CP_\pm=P_\mp^T C$
one derives
\begin{eqnarray}
-iI\Gamma_A \epsilon \thetab\, \bar \thetab \Gamma^{A}P_+
&=&
-iP_+IP_-\Gamma_{\bar A}P_+ \epsilon P_+ \thetab\, \bar \thetab P_- \Gamma^{\bar A}P_+~,\\
\frac{i}{2} \Gamma_{AB} \thetab\,\bar\thetab \hat \Gamma^{AB} \epsilon P_+
&=&
P_+\frac{i}{2} \Gamma_{\bar A\bar B} \thetab\,\bar\thetab P_- \hat \Gamma^{\bar A\bar B}  P_+\epsilon P_+
+P_+\frac{i}{2} \Gamma_{\underline A \underline B} \thetab\,\bar\thetab P_- \hat \Gamma^{\underline A \underline B} P_+
\epsilon P_+~,
\end{eqnarray}
so that
\begin{eqnarray}
\CM^2P_+ = P_+ \CM^2 P_+~.
\label{CM^2 2}
\end{eqnarray}
Gathering the results \bref{D theta 2} and \bref{CM^2 2} together we  obtain
$\bm L_\tau = P_+ \bm L_\tau$.
Using this we may derive
\begin{eqnarray}
\bar{\bm\lambda} I\sigma_1 \bm L_\tau
&=&
\bar{\bm\lambda} I\sigma_1 P_+\bm L_\tau
=\bar{\bm\lambda} P_+ IP_-\sigma_1 P_+\bm L_\tau
=\overline {P_- \bm\lambda} P_+ IP_-\sigma_1 P_+\bm L_\tau=0~.
\end{eqnarray}
It implies that the boundary condition for $p=3$ mod 4
eliminates the surface term \bref{full surface term}.

\bigskip

Summarizing we have shown that the boundary condition 
\bref{bc 1} with \bref{bc 2}
eliminates the surface term
\bref{surface 2} of the BRST transformation $\varepsilon Q(S)$.

\section{Summary and discussions}

We examined the BRST invariance of the open pure spinor superstring action
in the \adss{5}{5} background.
In order for the BRST symmetry to be preserved even in the presence of the boundary,
the surface term of the BRST transformation
must be eliminated by appropriate boundary conditions.
We determined such boundary conditions and 
found that the boundary conditions lead to a classification of possible
configurations of
1/2 supersymmetric D-branes in the \adss{5}{5} background.
Our result is summarized in the Table 1.
This is consistent with the results obtained by the other approaches \cite{Dp AdS pp,GS kappa 1,GS kappa 2}.
\medskip

We have used an exponential parametrization of the coset representative $g$
throughout this paper. 
In \cite{RS} a GS superstring action in the \adss{5}{5} background
  was derived based on an alternate version of the coset superspace construction in terms of GL(4$|$4).
The pure spinor superstring action
in this coset superspace construction was given in \cite{PS RS}.
This action is expected to make it more transparent
to examine the surface term for the BRST transformation of the action
and to derive possible
D-brane configurations  in the \adss{5}{5} background.

\medskip

The method used in this paper can be applied easily to
a superstring in 
the other background, 
for example the superstring in the type IIB pp-wave background  \cite{GS pp 1}\cite{PS IIB pp}.
The result will be consistent with the one obtained by the boundary $\kappa$-invariance of the  open 
GS superstring \cite{GS kappa 1,Bain}
and by examining equations of motion for a D-brane \cite{Dp AdS pp}.

It is also known that in the presence of a constant flux,
the boundary condition to ensure the $\kappa$-invariance
of the GS superstring action
leads to possible (non-commutative) D-branes
\cite{NC D}.
Furthermore the boundary condition to ensure the $\kappa$-invariance
of the supermembrane action
leads to the self-duality condition for
 the three-form flux on the M5-brane world-volume \cite{NC M}.
The same result is expected to be obtained by using
the pure spinor supermembrane action \cite{supermembrane PS}.
We hope to report this issue in another place \cite{HS2}.

\medskip

Finally let us  comment on a characterization of the Wess-Zumino (WZ) action.
The WZ action is necessary for the $\kappa$-symmetry of the action,
and then halves fermionic degrees of freedom  on the world-volume 
so as to match bosonic and fermionic degrees of freedom.
It is shown that the WZ term of a (D)$p$-brane
in flat spacetime is characterized
as a non-trivial element of the Chevalley-Eilenberg (CE) cohomology
in \cite{dAT} for $p$-branes,
and in \cite{CdAIB,MS} for D$p$-branes.
In \cite{Grassi}, D$p$-brane actions
in the extended  pure spinor formalism \cite{without}
are characterized as a non-trivial element of the BRST cohomology
of the extended BRST symmetry.
 It is interesting for us to extend this analysis to the  D$p$-brane action in
the  \adss{5}{5} background.\footnote{
On a CE cohomology classification
of D$p$-brane actions in the \adss{5}{5} background
see e.g.
\cite{CE AdS,HK}.}

\section*{Acknowledgments}

The authors would like to thank Takanori Fujiwara, Yoshifumi Hyakutake and Kentaroh Yoshida
 for useful comments.

\appendix

\section*{Appendix}

\section{Notation and convention}
The super-isometry  algebra of the \adss{5}{5}
background
is  $\mathfrak{psu}(2,2|4)$
of which (anti-)commutation relations are
\footnote{We follow the notation given in \cite{HKS0202}
except for $\Gamma_{11}$.}
\begin{eqnarray}
[P_a,P_b]&=&M_{ab}~,~~~[P_{a'},P_{b'}]=-M_{a'b'}~,\nn\\
{[}M_{AB},P_C]&=&\eta_{BC}P_A-\eta_{AC}P_B~,~~~
[M_{AB},M_{CD}]=\eta_{BC}M_{AD}+\mbox{3-terms}~,
\label{psu bosonic}
\end{eqnarray}
and
\begin{eqnarray}
{[}Q_I,M_{AB}]&=&-\frac{1}{2}Q_I\Gamma_{AB}~,~~~
[Q_I,P_A]=\frac{1}{2} \epsilon_{IJ}Q_J I\Gamma_A ~,~~~
\nn\\
\{Q_I,Q_J\}&=&-2iC\Gamma^A P_A \delta_{IJ} h_+
+\epsilon_{IJ}\left(
iC\Gamma^{ab} I M_{ab}
-iC\Gamma^{a'b'} I M_{a'b'}
\right)h_+,
\label{psu 32}
\end{eqnarray}
where
$\Gamma^A$ $(A=0,1,\cdots,9)$ are $32\times 32$ gamma matrices,
and $Q_I=Q_I h_+$ $(I=1,2)$ are a pair of 
Majorana-Weyl spinors.
We introduced $\epsilon_{IJ} =\left(\begin{array}{cc}0 & 1 \\-1 & 0\end{array}\right)$.
We have defined  $I=\Gamma^{01234}$ and $h_\pm=\frac{1}{2} (1\pm \Gamma_{11}) $
with $\Gamma_{11}\equiv \Gamma^{012\cdots9}$.
The charge conjugation matrix $C$ satisfies $C\Gamma_A =-\Gamma_A^T C$.
The AdS$_5$ isometry is generated by $P_a$ an $M_{ab}$ ($a,b=0,1,\cdots,4$),
while the S$^5$ isometry is by  $P_{a'}$ and $M_{a'b'}$ ($a',b'=5,6,\cdots,9$).

The left-invariant Cartan one-form $\bm L$
is defined by
\begin{eqnarray}
\bm L&=&g^{-1}\d g
=\bm L^AP_A +\frac{1}{2}\bm L^{AB} M_{AB}+ Q_I  \bm L^I 
\end{eqnarray} 
where $g\in$ PSU(2,2$|$4)/(SO(1,4)$\times$ SO(5)).
Parametrizing $g$ by $g=g_x(x) \e^{Q_I\thetab^I}$
and introducing the vielbein $e^A(x)$ and spin-connection $w^{AB}(x)$
by $g_x^{-1}\d g_x=e^A P_A +\frac{1}{2}w^{AB}M_{AB}$,
we obtain
\begin{eqnarray}
\bm L^A&=& e^A-2i \bar\thetab \Gamma^A 
\left(\frac{1}{2} +\frac{\CM^2}{4!}+\frac{\CM^4}{6!}+\cdots\right)
 D\thetab~,\\
\bm L^{AB}&=&
w^{AB}
+2i \bar\thetab\hat\Gamma^{AB}\epsilon
\left(\frac{1}{2} +\frac{\CM^2}{4!}+\frac{\CM^4}{6!}+\cdots\right) D\thetab~,\\
\bm L^I&=&\left(1+\frac{\CM^2}{3!} +\frac{\CM^4}{5!}+\cdots\right)D\thetab~,
\label{J^I 32}
\end{eqnarray}
where we have defined
$\bar\thetab \equiv \thetab^T C$, $\hat\Gamma^{AB}\equiv(\Gamma^{ab} I,-\Gamma^{a'b'}I)$
and
\begin{eqnarray}
D\thetab&=&
\d\thetab
+\frac{1}{2}e^A I\Gamma_A \epsilon \thetab
+\frac{1}{4}w^{AB}\Gamma_{AB}\thetab~,
\label{D theta 32}\\
\CM^2&=& -iI\Gamma_A \epsilon\thetab\cdot 
\bar\thetab\Gamma^A
+\frac{i}{2}\Gamma_{AB}\thetab\cdot\bar\thetab\hat\Gamma^{AB}\epsilon~.
\label{M^2 32}
\end{eqnarray}

\subsection{16-component spinor notation}

Let us rewrite \bref{psu 32} in terms of $16\times 16$ gamma-matrices $\gamma^A$.
We decompose $\Gamma^A$ as
\begin{eqnarray}
\Gamma^0=\1_{16}\otimes i\sigma_2~,~~
\Gamma^i=\gamma^i\otimes \sigma_1~,~~
\Gamma^9 = \gamma \otimes \sigma_1
\end{eqnarray}
where $\gamma\equiv  \gamma^{1\cdots 8}$ and $i=1,2,\cdots,8$.
It implies that
\footnote{
In the literature, $\gamma^A$ and $\tilde \gamma^A$ are distinguished each other by putting spinor indices.
In this paper, however, we use a matrix notation without spinor indices to make our presentation simpler.
}
\begin{eqnarray}
\Gamma^A=\left(\begin{array}{cc}0 & \tilde \gamma^A \\\gamma^A & 0\end{array}\right)~,~~
\tilde \gamma^A\equiv (1,\gamma^i,\gamma)~,~~
\gamma^A\equiv (-1,\gamma^i,\gamma)~.
\end{eqnarray}
The anti-commutation relation  $\{\Gamma^A,\Gamma^B\}=2\eta^{AB}\1_{32}$ turns out to
\begin{eqnarray}
\tilde \gamma^A\gamma^B +\tilde \gamma^B\gamma^A = 2\eta^{AB} \1_{16}~,~~~
\gamma^A\tilde \gamma^B +\gamma^B\tilde  \gamma^A = 2\eta^{AB} \1_{16}~.
\end{eqnarray}
It follows that
\begin{eqnarray}
\Gamma_{AB}&=&\left(\begin{array}{cc}\tilde\gamma_{AB} & 0 \\0 & \gamma_{AB}\end{array}\right)~,~~
I=\left(\begin{array}{cc}0 & \tilde \gamma^{0\cdots 4} \\ \gamma^{0\cdots 4} & 0\end{array}\right)~,
~~
I\Gamma_A=\left(\begin{array}{cc}\tilde \gamma^{0\cdots 4}\gamma_A & 0 \nn\\
0 & \gamma^{0\cdots 4}\tilde\gamma_A \end{array}\right)~,\nn\\
\Gamma_{11}&=&\left(\begin{array}{cc}\tilde\gamma^{0}\gamma^1 \cdots \gamma^{9} & 0 \\
0 & \gamma^{0}\tilde \gamma^1\cdots \gamma^9\end{array}\right)
=\left(\begin{array}{cc}1 & 0 \\0 & -1\end{array}\right)~,~~
C=\Gamma_0^\dag=\left(\begin{array}{cc}0 & 1 \\-1 & 0\end{array}\right)~,~~\nn\\
C\Gamma^A&=&\left(\begin{array}{cc}\gamma^A & 0 \\0 & -\tilde\gamma^A\end{array}\right)~,~~
C\Gamma^{AB}I=\left(\begin{array}{cc}\gamma^{AB}\gamma^{0\cdots 4} & 0 \\0 & -\tilde\gamma^{AB}\tilde \gamma^{0\cdots 4}\end{array}\right)~,
\end{eqnarray}
where we have defined the following objects
\begin{eqnarray}
&&\tilde\gamma_{AB}\equiv\frac{1}{2}(\tilde\gamma_A\gamma_B-\tilde\gamma_B\gamma_A)~,~~
\gamma_{AB}\equiv\frac{1}{2}(\gamma_A\tilde\gamma_B-\gamma_B\tilde\gamma_A)~,~~\nn\\
&&\tilde \gamma^{0\cdots 4}=\tilde\gamma^0\gamma^1\tilde\gamma^2\gamma^3\tilde\gamma^4~,~~
\gamma^{0\cdots 4}=\gamma^0\tilde\gamma^1\gamma^2\tilde\gamma^3\gamma^4~.
\end{eqnarray}
As $I^2=-1$, we have $ \gamma^{0\cdots 4}\tilde  \gamma^{0\cdots 4} = \tilde\gamma^{0\cdots 4} \gamma^{0\cdots 4}=-1$.
$Q_Ih_+=Q_I$ implies that $Q_I=(q_I,0)$
with $q_I$ ($I=1,2$) being a pair of 16-component spinors.
Similarly, $h_+\thetab=\thetab$ implies that $\thetab^I=\left(\begin{array}{c}\theta^I \\0\end{array}\right)$
with $\theta^I$ ($I=1,2$) being  a pair of 16-component spinors.
By using $q_I$, \bref{psu 32} takes of the form
\begin{eqnarray}
[q_I,M_{AB}]&=&-\frac{1}{2} q_I \tilde \gamma_{AB}~,~~
[q_I,P_A]=\frac{1}{2}\epsilon_{IJ} q_J \tilde\gamma^{0\cdots 4} \gamma_A~,\nn\\
\{q_I,q_J\}&=&-2i \gamma^A P_A \delta_{IJ} 
+\epsilon_{IJ}(i\gamma^{ab}\gamma^{0\cdots 4}M_{ab}-i\gamma^{a'b'}\gamma^{0\cdots 4}M_{a'b'})~.
\label{psu fermionic 16}
\end{eqnarray}
The fact that $C\Gamma^{A_1\cdots A_n}$ is symmetric iff $n=1,2 $ mod 4
implies that
 $\gamma_A$, $\tilde \gamma_A$, $\gamma_{A_1\cdots A_5}$
 and $\tilde\gamma_{A_1\cdots A_5}$ are symmetric
 and that $\gamma_{ABC}$ and  $\tilde\gamma_{ABC}$
 are antisymmetric.

In this notation,
the left-invariant Cartan one-forms are given as
\begin{eqnarray}
L&=&
L^AP_A +\frac{1}{2}L^{AB} M_{AB}+ q_I L^I ~,\\
L^A&=&
e^A-2i \theta\gamma^A
\left(\frac{1}{2}+\frac{m^2}{4!}+\frac{m^4}{6!}+\cdots\right) 
 D\theta~,\\
L^{AB}&=&w^{AB}
+2i\theta\hat\gamma^{AB}\epsilon
 \left(\frac{1}{2}+\frac{m^2}{4!}+\frac{m^4}{6!}+\cdots\right) 
 D\theta~,\\
 L^I&=& \left(1+\frac{m^2}{3!}+\frac{m^4}{5!}+\cdots\right) 
 D\theta~,
 \label{Cartan J^I}
\end{eqnarray} 
where
$\hat\gamma^{AB}=(\gamma^{ab}\gamma^{0\cdots 4},-\gamma^{a'b'}\gamma^{0\cdots 4})$
and
\begin{eqnarray}
D\theta&=&
\d\theta+\frac{1}{2}e^A\tilde\gamma^{0\cdots 4}\gamma_A \epsilon \theta
+\frac{1}{4}w^{AB}\tilde\gamma_{AB}\theta~,
\label{D thetra}\\
m^2&=&-i \tilde\gamma^{0\cdots 4} \gamma_A\epsilon\theta\cdot \theta\gamma^A
+\frac{i}{2}\tilde\gamma_{AB}\theta\cdot\theta\epsilon \hat \gamma^{AB}
~.
\label{m^2}
\end{eqnarray}

The currents used in \bref{J}
are related to the above objects by
\begin{eqnarray}
J_{0\xi}=\frac{1}{2} L_\xi^{AB}M_{AB}~,~~
J_{1\xi}=q_{\alpha}L_\xi^{1\alpha}~,~~
J_{2\xi} = L_\xi^AP_A~,~~
J_{3\xi}=\hat q_{\hat\alpha}L_\xi^{2\hat \alpha}~,~~
\label{J 16}
\end{eqnarray}
where we have replaced $(q^1,q^2)$ with  $(q_\alpha, \hat q_{\hat\alpha})$ 
and correspondingly $(\theta^1,\theta^2)$ with $(\theta^{\alpha}, \hat\theta^{\hat\alpha})$.



\begin{thebibliography}{9}

\bibitem{PS}
  N.~Berkovits,
  ``Super-Poincar\'e covariant quantization of the superstring,''
  JHEP {\bf 0004} (2000) 018
  \href{http://arxiv.org/abs/hep-th/0001035}{[hep-th/0001035]}.



\bibitem{without}
  P.~A.~Grassi, G.~Policastro, M.~Porrati and P.~Van Nieuwenhuizen,
  ``Towards covariant quantization of superstrings without pure spinor constraints,''
  JHEP {\bf 0210} (2002) 054
  \href{http://arxiv.org/abs/hep-th/0112162}{[hep-th/0112162]}.\\
  P.~A.~Grassi, G.~Policastro and P.~van Nieuwenhuizen,
  ``On the BRST Cohomology of Superstrings with/without Pure Spinors,''
  Adv.\ Theor.\ Math.\ Phys.\  {\bf 7} (2003) no.3,  499
  \href{http://arxiv.org/abs/hep-th/0206216}{[hep-th/0206216]}.

  \bibitem{without 2}
  Y.~Aisaka and Y.~Kazama,
  ``A New first class algebra, homological perturbation and extension of pure spinor formalism for superstring,''
  JHEP {\bf 0302} (2003) 017
  \href{http://arxiv.org/abs/hep-th/0212316}{[hep-th/0212316]};
  ``Operator mapping between RNS and extended pure spinor formalisms for superstring,''
  JHEP {\bf 0308} (2003) 047
  \href{http://arxiv.org/abs/hep-th/0305221}{[hep-th/0305221]}
  ;
  ``Relating Green-Schwarz and extended pure spinor formalisms by similarity transformation,''
  JHEP {\bf 0404} (2004) 070
  \href{http://arxiv.org/abs/hep-th/0404141}{[hep-th/0404141]}.

\bibitem{doubled}
  Y.~Aisaka and Y.~Kazama,
  ``Origin of pure spinor superstring,''
  JHEP {\bf 0505} (2005) 046
  \href{http://arxiv.org/abs/hep-th/0502208}{[hep-th/0502208]}.

  
\bibitem{PS quantum}
  N.~Berkovits,
  ``Quantum consistency of the superstring in \adss{5}{5}  background,''
  JHEP {\bf 0503} (2005) 041
  \href{http://arxiv.org/abs/hep-th/0411170}{[hep-th/0411170]}.
  
\bibitem{PS one-loop}
  B.~C.~Vallilo,
  ``One loop conformal invariance of the superstring in an \adss{5}{5} background,''
  JHEP {\bf 0212} (2002) 042
  \href{http://arxiv.org/abs/hep-th/0210064}{[hep-th/0210064]}.
  
  
\bibitem{GS AdS5xS5}
  R.~R.~Metsaev and A.~A.~Tseytlin,
  ``Type IIB superstring action in \adss{5}{5}  background,''
  Nucl.\ Phys.\ B {\bf 533} (1998) 109
  \href{http://arxiv.org/abs/hep-th/9805028}{[hep-th/9805028]}.
  
  
\bibitem{BPR}
  I.~Bena, J.~Polchinski and R.~Roiban,
  ``Hidden symmetries of the \adss{5}{5}  superstring,''
  Phys.\ Rev.\ D {\bf 69} (2004) 046002
  \href{http://arxiv.org/abs/hep-th/0305116}{[hep-th/0305116]}.
  
  
\bibitem{PS AdS integrable}
  B.~C.~Vallilo,
  ``Flat currents in the classical \adss{5}{5}  pure spinor superstring,''
  JHEP {\bf 0403} (2004) 037
  \href{http://arxiv.org/abs/hep-th/0307018}{[hep-th/0307018]}.\\
  I.~Adam, A.~Dekel, L.~Mazzucato and Y.~Oz,
  ``Integrability of Type II Superstrings on Ramond-Ramond Backgrounds in Various Dimensions,''
  JHEP {\bf 0706} (2007) 085
  \href{http://arxiv.org/abs/hep-th/0702083}{[hep-th/0702083 [HEP-TH]]}.

\bibitem{integrability GS RS}
  M.~Hatsuda and K.~Yoshida,
  ``Classical integrability and super Yangian of superstring on \adss{5}{5},''
  Adv.\ Theor.\ Math.\ Phys.\  {\bf 9} (2005) no.5,  703
  \href{http://arxiv.org/abs/hep-th/0407044}{[hep-th/0407044]}.
  
\bibitem{Maldacena}
  J.~M.~Maldacena,
  ``The Large N limit of superconformal field theories and supergravity,''
  Int.\ J.\ Theor.\ Phys.\  {\bf 38} (1999) 1113
   [Adv.\ Theor.\ Math.\ Phys.\  {\bf 2} (1998) 231]
  \href{http://arxiv.org/abs/hep-th/9711200}{[hep-th/9711200]}.\\
  S.~S.~Gubser, I.~R.~Klebanov and A.~M.~Polyakov,
  ``Gauge theory correlators from noncritical string theory,''
  Phys.\ Lett.\ B {\bf 428} (1998) 105
  \href{http://arxiv.org/abs/hep-th/9802109}{[hep-th/9802109]}.\\
  E.~Witten,
  ``Anti-de Sitter space and holography,''
  Adv.\ Theor.\ Math.\ Phys.\  {\bf 2} (1998) 253
  \href{http://arxiv.org/abs/hep-th/9802150}{[hep-th/9802150]}.
  



  
  
\bibitem{PS open background}
  N.~Berkovits and V.~Pershin,
  ``Supersymmetric Born-Infeld from the pure spinor formalism of the open superstring,''
  JHEP {\bf 0301} (2003) 023
  \href{http://arxiv.org/abs/hep-th/0205154}{[hep-th/0205154]}.
  
\bibitem{PS closed background}
  N.~Berkovits and P.~S.~Howe,
  ``Ten-dimensional supergravity constraints from the pure spinor formalism for the superstring,''
  Nucl.\ Phys.\ B {\bf 635} (2002) 75
  \href{http://arxiv.org/abs/hep-th/0112160}{[hep-th/0112160]}.
  
  
  
\bibitem{Dp AdS pp}
  K.~Skenderis and M.~Taylor,
  ``Branes in AdS and pp wave space-times,''
  JHEP {\bf 0206} (2002) 025
  \href{http://arxiv.org/abs/hep-th/0204054}{[hep-th/0204054]}.
    

  
\bibitem{GS kappa 1}
  M.~Sakaguchi and K.~Yoshida,
  ``D-branes of covariant AdS superstrings,''
  Nucl.\ Phys.\ B {\bf 684} (2004) 100
  \href{http://arxiv.org/abs/hep-th/0310228}{[hep-th/0310228]}. 



 \bibitem{AdS PS action}
  N.~Berkovits and O.~Chandia,
  ``Superstring vertex operators in an \adss{5}{5}  background,''
  Nucl.\ Phys.\ B {\bf 596} (2001) 185
  \href{http://arxiv.org/abs/hep-th/0009168}{[hep-th/0009168]}.
  


\bibitem{review}  
  N.~Berkovits,
  ``ICTP lectures on covariant quantization of the superstring,''
  \href{http://arxiv.org/abs/hep-th/0209059}{[hep-th/0209059]}.\\
  Y.~Oz,
  ``The pure spinor formulation of superstrings,''
  Class.\ Quant.\ Grav.\  {\bf 25} (2008) 214001
  \href{http://arxiv.org/abs/0910.1195}{[arXiv:0910.1195 [hep-th]]}.\\
  O.~A.~Bedoya and N.~Berkovits,
  ``GGI Lectures on the Pure Spinor Formalism of the Superstring,''
  \href{http://arxiv.org/abs/0910.2254}{[arXiv:0910.2254 [hep-th]]}.\\
   L.~Mazzucato,
  ``Superstrings in AdS,''
  Phys.\ Rept.\  {\bf 521} (2012) 1
  \href{http://arxiv.org/abs/1104.2604}{[arXiv:1104.2604 [hep-th]]}.
  

 
\bibitem{GS kappa 2}
  M.~Sakaguchi and K.~Yoshida,
  ``Notes on D-branes of type IIB string on \adss{5}{5} ,''
  Phys.\ Lett.\ B {\bf 591} (2004) 318
  \href{http://arxiv.org/abs/hep-th/0403243}{[hep-th/0403243]}.

\bibitem{LW}
  N.~D.~Lambert and P.~C.~West,
  ``D-branes in the Green-Schwarz formalism,''
  Phys.\ Lett.\ B {\bf 459} (1999) 515
  \href{http://arxiv.org/abs/hep-th/9905031}{[hep-th/9905031]}.

\bibitem{Bain}
  P.~Bain, K.~Peeters and M.~Zamaklar,
  ``D-branes in a plane wave from covariant open strings,''
  Phys.\ Rev.\ D {\bf 67} (2003) 066001
  \href{http://arxiv.org/abs/hep-th/0208038}{[hep-th/0208038]}.



\bibitem{RS}
  R.~Roiban and W.~Siegel,
  ``Superstrings on \adss{5}{5}  supertwistor space,''
  JHEP {\bf 0011} (2000) 024
  \href{http://arxiv.org/abs/hep-th/0010104}{[hep-th/0010104]}.
    
    \bibitem{PS RS}
  I.~Ramirez and B.~C.~Vallilo,
 ``Supertwistor description of the AdS pure spinor string,''
  Phys.\ Rev.\ D {\bf 93} (2016) no.8,  086008
  \href{http://arxiv.org/abs/1510.08823}{[arXiv:1510.08823 [hep-th]]}.



\bibitem{GS pp 1}
  R.~R.~Metsaev,
  ``Type IIB Green-Schwarz superstring in plane wave Ramond-Ramond background,''
  Nucl.\ Phys.\ B {\bf 625} (2002) 70
  \href{http://arxiv.org/abs/hep-th/0112044}{[hep-th/0112044]}.\\
  R.~R.~Metsaev and A.~A.~Tseytlin,
  ``Exactly solvable model of superstring in Ramond-Ramond plane wave background,''
  Phys.\ Rev.\ D {\bf 65} (2002) 126004
  \href{http://arxiv.org/abs/hep-th/0202109}{[hep-th/0202109]}.

    
\bibitem{PS IIB pp}
  N.~Berkovits,
  ``Conformal field theory for the superstring in a Ramond-Ramond plane wave background,''
  JHEP {\bf 0204} (2002) 037
  \href{http://arxiv.org/abs/hep-th/0203248}{[hep-th/0203248]}.

  
  
\bibitem{NC D}
  M.~Sakaguchi and K.~Yoshida,
  ``Noncommutative D-brane from Covariant AdS Superstring,''
  Nucl.\ Phys.\ B {\bf 797} (2008) 179
  \href{http://arxiv.org/abs/hep-th/0604039}{[hep-th/0604039]}.
  
\bibitem{NC M}
  M.~Sakaguchi and K.~Yoshida,
  ``Noncommutative M-branes from covariant open supermembranes,''
  Phys.\ Lett.\ B {\bf 642} (2006) 400
  \href{http://arxiv.org/abs/hep-th/0608099}{[hep-th/0608099]}
  ;
  ``Intersecting Noncommutative M5-branes from Covariant Open Supermembrane,''
  Nucl.\ Phys.\ B {\bf 781} (2007) 85
  \href{http://arxiv.org/abs/hep-th/0702062}{[hep-th/0702062 [HEP-TH]]}
;
  ``A Covariant Approach to Noncommutative M5-branes,''
  Prog.\ Theor.\ Phys.\ Suppl.\  {\bf 171} (2007) 275
  \href{http://arxiv.org/abs/hep-th/0702132}{[hep-th/0702132 [HEP-TH]]}.


\bibitem{supermembrane PS}
  N.~Berkovits,
  ``Towards covariant quantization of the supermembrane,''
  JHEP {\bf 0209} (2002) 051
  \href{http://arxiv.org/abs/hep-th/0201151}{[hep-th/0201151]}.
  


\bibitem{HS2}
S.~Hanazawa and M.~Sakaguchi, work in progress.
  
\bibitem{dAT}
  J.~A.~De Azcarraga and P.~K.~Townsend,
  ``Superspace Geometry and Classification of Supersymmetric Extended Objects,''
  Phys.\ Rev.\ Lett.\  {\bf 62} (1989) 2579.

\bibitem{CdAIB}
  C.~Chryssomalakos, J.~A.~de Azcarraga, J.~M.~Izquierdo and J.~C.~Perez Bueno,
  ``The Geometry of branes and extended superspaces,''
  Nucl.\ Phys.\ B {\bf 567} (2000) 293
  [hep-th/9904137].

\bibitem{MS}
  M.~Sakaguchi,
  ``IIB Branes and new space-time superalgebras,''
  JHEP {\bf 0004} (2000) 019
  [hep-th/9909143].
  
  
  \bibitem{Grassi}
  L.~Anguelova and P.~A.~Grassi,
  ``Super D-branes from BRST symmetry,''
  JHEP {\bf 0311} (2003) 010
  [hep-th/0307260].
  

\bibitem{CE AdS}
  M.~Sakaguchi and K.~Yoshida,
  ``Non-relativistic AdS branes and Newton-Hooke superalgebra,''
  JHEP {\bf 0610} (2006) 078
  [hep-th/0605124].

  \bibitem{HK}
  M.~Hatsuda and K.~Kamimura,
  ``Wess-Zumino terms for AdS D-branes,''
  Nucl.\ Phys.\ B {\bf 703} (2004) 277
  [hep-th/0405202].



\bibitem{HKS0202}
  M.~Hatsuda, K.~Kamimura and M.~Sakaguchi,
``From Super \adss{5}{5} algebra to super pp wave algebra,''
  Nucl.\ Phys.\ B {\bf 632} (2002) 114
  \href{http://arxiv.org/abs/hep-th/0202190}{[hep-th/0202190]}.
  
  
\end{thebibliography}
\end{document}